\documentclass[10pt,journal,compsoc]{IEEEtran}

\usepackage[utf8]{inputenc}
\usepackage{enumitem}

\usepackage[table]{xcolor}
\usepackage{xcolor}

\colorlet{shadecolor}{yellow}
\usepackage[pdftex]{graphicx}
\graphicspath{{figure_transaction/}}
\DeclareGraphicsExtensions{.pdf,.jpeg,.png,.eps}

\usepackage[cmex10]{amsmath}
\usepackage{array}
\usepackage{mdwmath}
\usepackage{mdwtab}
\usepackage{eqparbox}
\usepackage{url}
\usepackage{longtable}
\usepackage{lipsum}
\usepackage{blindtext}
\usepackage{amsmath}
\usepackage{multirow}
\usepackage[flushleft]{threeparttable}
\usepackage{color}
\usepackage{times}
\usepackage{epsfig}
\usepackage{graphicx}
\usepackage{amsmath}
\usepackage{amssymb}
\usepackage{tabularx}
\usepackage{cite}
\usepackage{makecell}
\usepackage{comment}
\usepackage[normalem]{ulem}
\usepackage{tabularx}
\usepackage{import}
\usepackage{array}
\usepackage{boldline}
\usepackage{caption}
\usepackage{subcaption}
\def \hfillx {\hspace*{-\textwidth} \hfill}
\def \vfillx {\vspace*{10pt} \vfill}
\usepackage{hyperref}
\hypersetup{
    colorlinks=true,
    linkcolor=black,
    filecolor=black,      
    urlcolor=black,
}

\usepackage[ruled,vlined,linesnumbered]{algorithm2e}
\newcolumntype{C}[1]{>{\centering\arraybackslash}m{#1}}
\newcolumntype{C}{>{\hsize=\dimexpr2\hsize+8.25\tabcolsep+\arrayrulewidth\centering\relax}X}
\newcolumntype{U}{>{\hsize=\dimexpr2\hsize+6\tabcolsep+\arrayrulewidth\centering\relax}X}
\newcolumntype{Q}{>{\hsize=\dimexpr1\hsize+1\tabcolsep+\arrayrulewidth\centering\relax}X}
\newcolumntype{S}{>{\hsize=\dimexpr1\hsize+1.1\tabcolsep+\arrayrulewidth\centering\relax}X}
\newcolumntype{O}{>{\hsize=\dimexpr1\hsize+18\tabcolsep+\arrayrulewidth\centering\relax}X}
\newcolumntype{P}{>{\hsize=\dimexpr1\hsize+9\tabcolsep+\arrayrulewidth\centering\relax}X}

\newcolumntype{Y}{>{\centering\arraybackslash}X}
\ifCLASSINFOpdf
\else
\fi
\begin{document}

\title{Self-supervised ECG Representation Learning for Emotion Recognition}

\author{Pritam Sarkar, Ali Etemad \\ \textit{\{pritam.sarkar, ali.etemad\}@queensu.ca} \\ Department of Electrical and Computer Engineering,\\ Queen's University, Kingston, Canada

\thanks{
2020 IEEE. Personal use of this material is permitted. Permission from IEEE must be obtained for all other uses, in any current or future media, including reprinting/republishing this material for advertising or promotional purposes, creating new collective works, for resale or redistribution to servers or lists, or reuse of any copyrighted component of this work in other works
}
}

\IEEEtitleabstractindextext{%
\begin{abstract}
We exploit a self-supervised deep multi-task learning framework for electrocardiogram (ECG) -based emotion recognition. The proposed solution consists of two stages of learning \textit{a)} learning ECG representations and \textit{b)} learning to classify emotions. ECG representations are learned by a signal transformation recognition network. The network learns high-level abstract representations from unlabeled ECG data. Six different signal transformations are applied to the ECG signals, and transformation recognition is performed as pretext tasks. Training the model on pretext tasks helps the network learn spatiotemporal representations that generalize well across different datasets and different emotion categories. We transfer the weights of the self-supervised network to an emotion recognition network, where the convolutional layers are kept frozen and the dense layers are trained with labelled ECG data. We show that the proposed solution considerably improves the performance compared to a network trained using fully-supervised learning. New state-of-the-art results are set in classification of arousal, valence, affective states, and stress for the four utilized datasets. Extensive experiments are performed, providing interesting insights into the impact of using a multi-task self-supervised structure instead of a single-task model, as well as the optimum level of difficulty required for the pretext self-supervised tasks. 
\end{abstract}

}

\maketitle

\section{Introduction}
Affective computing is a field of study that deals with understanding human emotions, intelligent human-machine interaction, and computer-assisted learning among others \cite{picard2000affective, picard2001toward}. The goal of affective computing is to equip machines with the ability to model and interpret the emotional states of humans. Emotion is considered a \textit{physiological} and \textit{psychological} expression associated with moods and personalities of individuals. As a result, sensing technologies integrated into wearable devices coupled with machine learning and deep learning techniques have recently been used to analyze physiological signals in order to classify or quantify human emotions \cite{pritam_acii, amigos_dataset, koelstra2011deap}. Recent applications of affective computing include human monitoring systems \cite{taylor2017personalized, shashikumar2017deep}, stress or anxiety level detection \cite{healey2005detecting, swell_dataset}, emotion and personality recognition \cite{amigos_dataset, subramanian2016ascertain}, and others. 

A wide variety of data sources have been utilized for affective computing purposes. These data sources include facial expressions \cite{sepas2019deep}, motion and gait \cite{etemad2014classification}, speech \cite{kwon2003emotion}, electrocardiogram (ECG) \cite{agrafioti2011ecg}, electroencephalogram (EEG) \cite{koelstra2011deap}, electrooculogram (EOG) \cite{zhang2019capsule}, and galvanic skin response (GSR) \cite{zhai2006stress} among others. In particular, ECG, which is the focus of our study, has been explored in the past with works such as \cite{agrafioti2011ecg} showing a strong correlation between emotional attributes and the ECG waveform, and investigating the feasibility and limitations of using ECG signals for emotion detection. Further, recent advancements in machine learning and deep learning have proven ECG to be a reliable source of information for emotion recognition systems \cite{pritam_acii, siddharth2019utilizing}.

The majority of machine learning or deep learning solutions for ECG-based emotion recognition utilize fully-supervised learning methods. Several limitations can be associated with this approach. First and foremost, in a typical setup of fully-supervised learning, the model needs to be trained from scratch for every classification or regression task \cite{nardelli2015recognizing}, which requires considerable computational resources and time. Additionally, the learned representations from the trained fully supervised models are often very task-specific, which are not expected to generalize well for other tasks. Lastly, fully supervised learning generally requires training using large human-annotated datasets, since small datasets typically result in poor performance with deeper networks.

\begin{figure}[t]
    \centering
    \includegraphics[width=1\columnwidth]{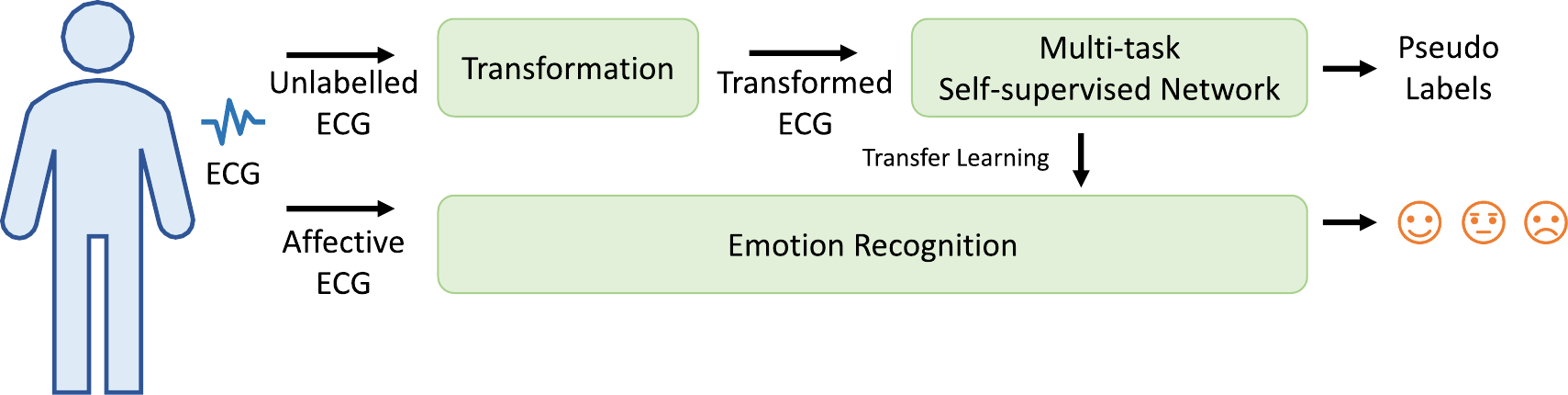}
    \caption{An overview of the proposed framework for self-supervised emotion recognition is presented.}
    \label{fig:framework}
\end{figure}

To tackle these problems in the context of ECG-based affective computing, inspired by the success of multi-task self-supervised learning in other domains notably \cite{multitask_selfsupervised}, we use a deep learning approach based on self-supervised representation learning \cite{raina2007self} (please see Figure \ref{fig:framework}). In self-supervised learning, models are trained using automatically generated labels instead of human-annotated ones. There are a number of advantages to self-supervised learning. First, the representations learned using this approach are often invariant to inter- and intra-instance variations \cite{wang2017transitive} due to learning rather generalized features as opposed to task-specific ones. As a result, these models can be reused for different tasks within the same domain. This property often improves the performance of the networks and also saves computation time compared to training a model from scratch for each task. Lastly, as self-supervised learning doesn't require human-annotated labels, larger datasets can be acquired and utilized, which in turn results in the ability to train deeper and more sophisticated networks. We use four publicly available datasets, AMIGOS \cite{amigos_dataset}, DREAMER \cite{dreamer_dataset}, WESAD \cite{wesad}, and SWELL \cite{swell_dataset} to perform emotion recognition with ECG. First, a number of transformations are applied to the ECG signals. The automatically generated labels associated with the transformations are then utilized to carry out self-supervised learning of ECG representations through a deep multitask Convolutional Neural Network (CNN). Next, the convolutional layers responsible for learning the ECG representations are frozen and transferred to a second network. The dense layers of this new network are then trained with the labeled ECG data to perform emotion classification. 

Our contributions can be summarized as follows:
\begin{itemize}[leftmargin = 15pt]
    \item We exploit a self-supervised framework for emotion recognition based on multi-task ECG representation learning. While self-supervised techniques have been used for time-series representation learning in other domains (see \cite{multitask_selfsupervised, 9053569, tagliasacchi2019self}), we propose its use in ECG-based affective computing for the first time. The results show that the self-supervised network outperforms the same network for emotion recognition when trained in a fully-supervised fashion.
    \item We perform very thorough analyses on the parameters associated with the self-supervised learning tasks, providing interesting insights into the impact of different self-supervised tasks and their contribution towards learning of affective ECG representations. Additionally, we show that self-supervised multi-task learning results in better learned representations compared to single-task self-supervised learning. Our analysis also provides interesting insights into the relationship between the difficulty of self-supervised tasks and the original emotion recognition task, where there seems to be an optimum level of difficulty in which the network can learn strong representations in a self-supervised manner. The analysis shows that simpler tasks do not result in proper learning of ECG representations, whereas extremely hard tasks also prevent the network from learning generalized representations.
    \item We set new state-of-the-art results for all the undertaken emotion classification tasks namely arousal, valence, affective states, and stress recognition in the four utilized datasets AMIGOS, DREAMER, WESAD, and SWELL. We demonstrate that the ECG representations learned by the self-supervised model generalize very well across all four ECG datasets, consistently resulting in accurate emotion recognition. 
\end{itemize}

This paper is an extension of our work \cite{sarkar2019self}, compared to which this paper additionally includes the following: \textit{a)} Two additional publicly available datasets DREAMER and WESAD are used in this paper to further investigate the benefits of the self-supervised learning framework compared to fully-supervised methods; \textit{b)} The impact of different transformation tasks on the intended emotion recognition tasks are investigated in depth and presented in this paper; \textit{c)} The relationship between the difficulty of self-supervised tasks and emotion recognition is investigated; \textit{d)} The benefits of the multi-task architecture for the self-supervised transformation recognition network are compared to a single-task approach, and the results are presented; \textit{e)} We analyze and discuss the impact of different parameters associated with each of the signal transformations with respect to the network performance; \textit{f)} An analysis of the relationship between the learned representations and the depth of the proposed network is presented; \textit{g)} Lastly, the effect of utilizing an aggregation of multiple datasets for self-supervised training as opposed to individual datasets is investigated and presented. 

The rest of the paper is organized as follows. Section \ref{Background} discusses the background and related works on emotion recognition with ECG. We describe the methodology in Section \ref{Methodology}. The datasets used in this paper and the experiment setup are described in Section \ref{setup}. Section \ref{eval} presents the performance and results of the proposed architecture. A thorough analysis and discussion on the limitations of the framework is presented in Section \ref{Discussion}. Section \ref{conclusion_future_work} presents the summary and concluding remarks.

\section{Background and Related Work} \label{Background}

\subsection{Electrocardiography}
ECG signals contain information regarding cardiac electrical activity over a period of time, where an ECG beat is composed of a number of prominent waves: the \textit{P} wave, the \textit{QRS} complex, the \textit{T} wave \cite{agrafioti2011ecg},  and often the \textit{U} wave \cite{GOLDBERGER20067}. Each waveform contains significant information that can be used to understand the cardiac state of individuals. To acquire ECG signals, electrodes are placed on the surface of the skin. The most common sensor configurations include $12$-lead, $5$-lead, $3$-lead, or single-channel systems. ECG signals have been widely used to develop deep learning solutions, for example arrhythmia detection \cite{andrewng_nature}, emotion or stress detection \cite{agrafioti2011ecg, hsu2017automatic}, biometric identification \cite{page2015utilizing}, and others. Additionally, research has shown strong correlations between cardiovascular activity and emotions, as emotions can influence the autonomic nervous system which controls the rhythm of heart beats \cite{agrafioti2011ecg}. To extract useful information from heart beats, different feature extraction techniques have been utilized \cite{xu2009method, healey2005detecting}. Most common approaches include either utilizing raw ECG signals to directly learn high-level representations \cite{andrewng_nature}, or extracting time-domain and frequency domain features successive to detection of fiducial points (P, Q, R, S, and T components) from the ECG waveform \cite{agrafioti2011ecg, nardelli2015recognizing}. Successive to learning of ECG representations, classifiers such as Multi-Layer Perceptrons are typically used learn the relationships between the high-level representations and the main task \cite{pritam_acii, santamaria2018using}.

\subsection{ECG-based Affective Computing}
Systems capable of performing affective computing generally aim to classify or quantify the emotional state of individuals based on collected data and information. Recent advancements in affective computing have shown ECG to be a reliable source of information to classify the emotional states of humans \cite{subramanian2016ascertain, pritam_acii, sensor_kyle, siddharth2019utilizing}. The type of affective states often studied include stress, happiness, sorrow, and excitement among others. 

In an early work on ECG-based affective computing, stress detection was performed in \cite{healey2005detecting}, as participants went through a driving task. ECG signals were collected during the study, followed by extraction of time and frequency domain features which were then used to perform classification. In particular, Heart Rate Variability (HRV) features were calculated from power spectrum of ECG signals. The extracted features were used to perform classification using a linear discriminant classifier. 

In a large study, in \cite{xu2009method}, emotion recognition was performed, where ECG signals from $391$ subjects were captured, while individuals were shown a movie clip belonging to either the category of joy or sadness. Out of all the participants, $150$ subjects were selected who had shown effective elicitation during this study. Next, continuous wavelet transform was performed to accurately detect P-QRS-T wave locations. A total of $79$ features were extracted. To perform better and effective classification, feature selection was performed by means of a hybrid particle swarm optimization technique. Their study shows that the selected features can correctly classify emotions with a Fisher classifier, reporting an accuracy of $88.43\%$.

The relationship between personality and affect was studied in \cite{subramanian2016ascertain}. The study was conducted on $58$ participants. While subjects were shown emotional movie clips, ECG data were recorded. Similar to \cite{healey2005detecting}, HRV features were calculated, along with heart rate and inter-beat intervals. These features were utilized to perform classification of arousal and valence scores. Classification was performed by a Support Vector Machine (SVM) and Naive Bayes classifier. The naive Bayes classifier outperformed the SVM, reporting an accuracy of $59\%$ and $60\%$ for arousal and valence respectively.

In a recent work, deep learning was used to classify cognitive load and expertise of learners in a medical simulation using ECG \cite{pritam_acii}. ECG data were recorded from $9$ medical practitioners in $2$ classes, expert and novice. First, baseline ECG was collected as participants were in rest. Next, ECG signals were collected during simulation where the subjects had to tend to an injured manikin. R-peaks were then detected, followed by the extraction of $11$ time-domain features. Additionally, $9$ frequency domain features were extracted successive to calculating the power of different frequency bands. Finally, the extracted features were normalized by corresponding baseline features. A deep neural network with $7$ layers was employed to perform classification of expertise and cognitive load. The model achieved an accuracy of $89.4\%$ and $96.6\%$ for cognitive load and expertise respectively. The research shows an interesting relationships between level of expertise and cognitive load, indicating that under stressful conditions, experts tend show lower cognitive load compared to novice subjects. 

A number of prior works such as \cite{amigos_dataset, santamaria2018using, dreamer_dataset, siddharth2019utilizing, wesad}, and \cite{swell_dataset} have used one or more of the four datasets utilized in this paper, AMIGOS, DREAMER, WESAD, and SWELL, for emotion recognition. In \cite{amigos_dataset}, a Gaussian Naive Bayes (GNB) classifier was used on AMIGOS and reported F1 scores of $54.5\%$ and $55.1\%$ for classifying arousal and valence respectively. To perform a similar study, a CNN was used in \cite{santamaria2018using}, reporting accuracies of $81\%$ and $71\%$ for arousal and valence classification. An initial study on DREAMER was performed in \cite{dreamer_dataset} where arousal and valence classification was performed using an SVM classifier and an accuracy of $62.37\%$ was reported. A novel technique was employed in \cite{siddharth2019utilizing} to perform emotion recognition on both DREAMER and AMIGOS. Two different sets of features were fused to perform the classification. First HRV features were extracted by calculating RR intervals from the input ECG signals. Next, spectrogram images were generated from time-series ECG signals. A pre-trained VGG-$16$ was used to extract feature embeddings of size $4096$, followed by dimensionality reduction to $30$ dimensions. Finally, both sets of features were concatenated and used to perform classification by means of an Extreme Learning Machine (ELM) classifier. Three-level affect state detection was performed in \cite{wesad}, where a Linear Discriminant Analysis (LDA) classifier was used on WESAD and reported an accuracy of $66.29\%$. Later, \cite{lin2019explainable} proposed a Mutlimodal-Multisensory Sequential Fusion (MMSF) model, which showed that the proposed model can be trained on different signals with different sampling rates in the same training batch. When utilizing only ECG, an accuracy of $83\%$ was reported for detection of affect state on WESAD. In \cite{sriramprakash2017stress}, stress detection was performed in SWELL. Feature analysis was performed on extracted time domain and frequency domain features. Time domain feature NN$50$ showed the best performance compared to other features and reported an accuracy of $86.36\%$ by utilizing an SVM classifier.

\subsection{Self-supervised Representation Learning}
Recent research in the area of machine learning and deep learning has shown the potential of self-supervised models in learning generalized and robust representations. This concept has been utilized in computer vision \cite{wang2017transitive, kocabas2019self, wang20193d, 8058202, Wang_2019_CVPR}, speech \cite{tagliasacchi2019self}, natural language processing \cite{wu2010open}, and others. Self-supervised learning is a machine learning paradigm in which a network is trained using automatically generated labels instead of human-annotated labels. Different techniques have been employed to generate automatic labels. For example, a self-supervised network was proposed in \cite{jing2018self}, where high-level spatiotemporal features were learned from unlabelled videos. The network was trained using different rotated video clips to predict rotations. Next, the trained model was fine-tuned on the action recognition datasets. Their research showed that activity recognition accuracy was improved by more than $15\%$ on both utilized datasets, compared to a network trained in a fully-supervised manner. In another work in \cite{kocabas2019self}, $3$D pose estimation was tackled using self-supervised learning. To overcome the need for large amounts of $3$D ground truth data generally required for $3$D pose estimation, epipolar geometry was employed to calculate $3$D poses from the available $2$D poses. Subsequently, a convolution neural network was trained using the available $3$D poses and reported state-of-the-art results. Self-supervised learning has also been used in learning wearable data. For example, human activity recognition from Inertial Measurement Units (IMU) was performed in \cite{multitask_selfsupervised}. A set of different signal transformations were performed to train a convolutional neural network to predict transformations. Further, features were extracted from the trained network, and fully-connected layers were utilized to learn different activities such as walking, sitting, and jogging among others using the learned representations. 

The works presented above, among others, demonstrate the advantages and promising results obtained using self-supervised learning in other domains. Nonetheless, such a framework has not been proposed for ECG representation learning, which we undertake in this paper in the context of emotion recognition.

\section{Self-supervised Solution} \label{Methodology}
\begin{figure*}[t!]
    \centering
    \includegraphics[width=1\textwidth]{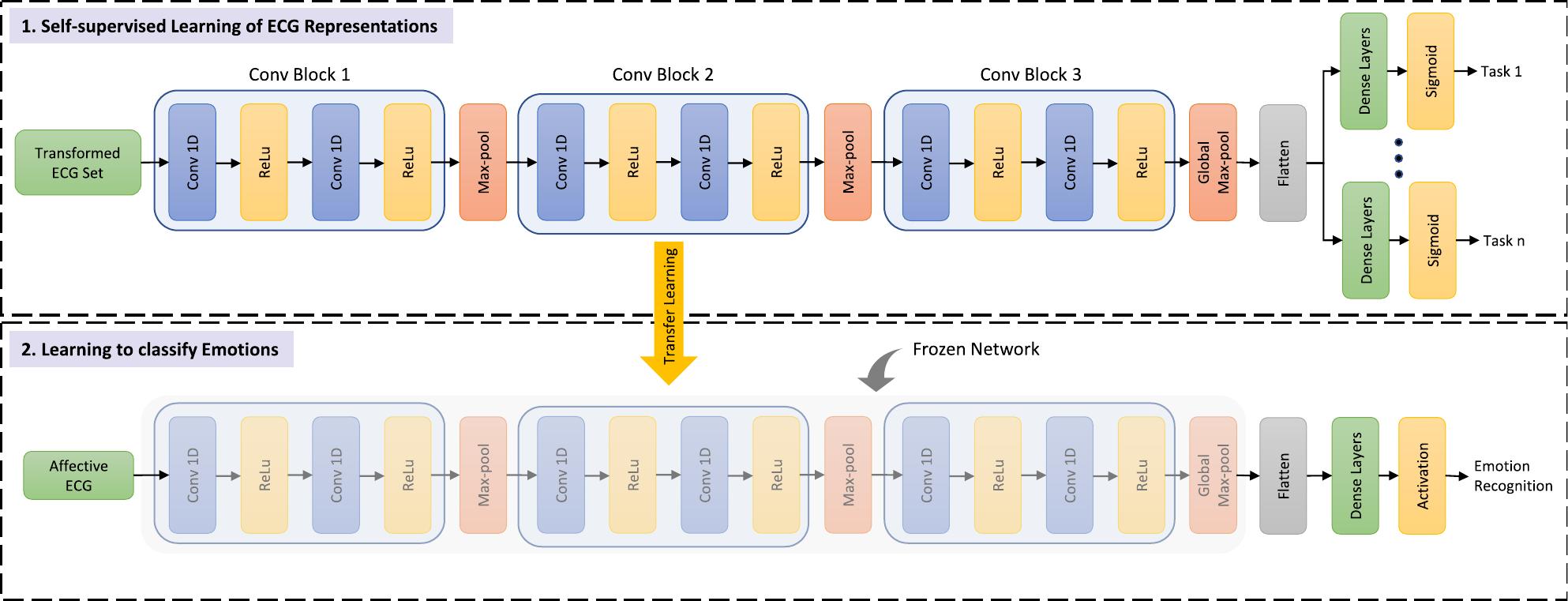}
    \caption{The self-supervised architecture is presented. First, a multi-task CNN is trained using automatically generated labels to learn ECG representations. Then, the weights are transferred to the emotion recognition network, where the fully connected layers are trained to classify emotions.}
    \label{fig:architecture}
\end{figure*}

Our goal in this paper is to learn ECG representations $R_{ECG}$ capable of distinguishing between different classes of emotion. The proposed framework consists of two stages of learning: first, learning ECG representation, and second, using the network capable of extracting ECG representation to learn another network capable of classifying emotions. Accordingly, we define two sets of tasks, \textit{pretext} and \textit{downstream}, referred to as $T_{p}$ and $T_{d}$ respectively, where $T_{p}$ is the set of signal transformation recognition tasks through which ECG representations are learned, while $T_{d}$ learns to classify emotions. The goal of $T_{p}$ is to learn robust generalized features from unlabelled ECG data through a self-supervised process, i.e. recognition of signal transformations. In the second stage in which $T_{d}$ is learned, the original ECG data and the human-annotated emotion labels $y_i$ are used. In this stage, the framework uses the frozen convolution layers of the first network to learn emotional classes upon supervised learning of the fully-connected layers at the end of the network. Figure \ref{fig:architecture} shows the proposed self-supervised solution. In the following subsections, we further describe the architecture and details of each learning stage. The optimum architecture and hyperparameters for these two networks described below are selected either empirically or based on searching among a large number of combinations for optimum performance. 

\begin{table}[]
    \footnotesize
    \centering
    \caption{The architecture of the signal transformation recognition network is presented.}
    \begin{tabular}{l|l|l}
    \hline
     \textbf{Module} & \textbf{Layer Details} & \textbf{Feature Shape} \\ \hline \hline
      Input & $-$ & $2560 \times 1$ \\ \hline
      \multirow{7}{*}{Shared Layers} & $[\textit{conv}, 1 \times 32, 32] \times 2$ &   $2560 \times 32$  \\ 
                                     &  $[\textit{maxpool}, 1 \times 8, \textit{stride} = 2]$ &   $1277 \times 32$  \\ 
                                     &  $[\textit{conv}, 1 \times 16, 64] \times 2$ &   $1277 \times 64$  \\ 
                                     &  $[\textit{maxpool}, 1 \times 8, \textit{stride} = 2]$ &   $635 \times 64$  \\ 
                                     &  $[\textit{conv}, 1 \times 8, 128] \times 2$ &   $635 \times 128$ \\ \cline{2-3} 
                                     & \textit{global max pooling} &  $1 \times 128$ \\ \hline 
      {\makecell[l]{Task-Specific\\Layers}} & \makecell[l]{$[\textit{dense}] \times 2$\\ $\times$ \textit{$7$ parallel tasks}} & 128  \\ \hline 
      Output & $-$ & $2$ \\ \hline

    \end{tabular}
    \label{tab:SSL_arch}
\end{table}

\subsection{Self-supervised Learning of ECG Representations}
We propose the use of a self-supervised deep multi-task CNN to learn $R_{ECG}$, through recognition of different transformations applied to input ECG signals (without utilizing any of human-annotated labels provided with the data). A similar approach was used in \cite{multitask_selfsupervised} for human activity recognition. Let's denote an example tuple of inputs and pseudo-labels for $T_{p}$ as $(X{_j}, P{_j})$ where $X{_j}$ is $j^{th}$ transformed signal, $P{_j}$ is the automatically generated label according to the $j^{th}$ transformation, and $j\in[0,N]$ where $N$ is the total number of signal transformations. To learn $T_{p}$, a signal transformation recognition network, $\gamma(X_j, P{_j}, \theta)$ is trained, where $\theta$ is the set of trainable parameters. The model learns $\theta$ by minimizing the total loss, $L_{total}$, composed of the weighted average of the individual losses of the signal transformation network ${L}_{j}$. The total network loss is defined by:
\begin{equation}
    \begin{aligned}
        {L}_{j} = [P_j \log \psi_j + (1 - P_j) \log (1 - \psi_j)],
    \end{aligned}
    \label{equ:individual_loss}
\end{equation}
\begin{equation}
    \begin{aligned}
        {L}_{total} = \sum_{j=0}^{N}{ {\alpha}_{j} {L}_{j}},
    \end{aligned}
    \label{equ:total_loss}
\end{equation}
where $\psi_j$ and ${\alpha}_{j}$ are the predicted probability and loss coefficient of the $j^{th}$ transformation task respectively.

\begin{figure}[t]
    \centering
    \includegraphics[width=1\columnwidth]{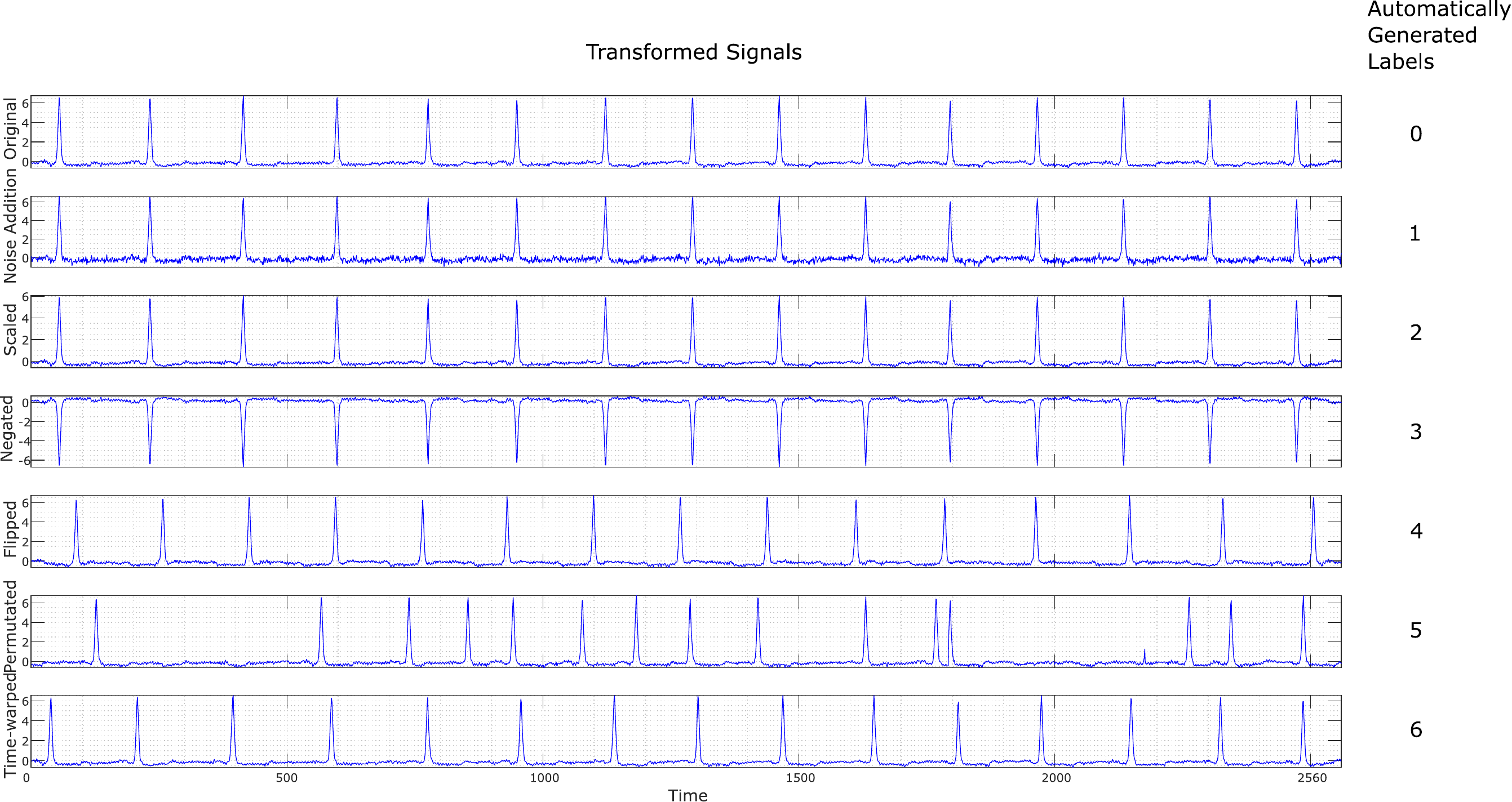}
    \caption[c]{A sample of an original ECG signal with the six transformed signals along with automatically generated labels are presented.}
    \label{fig:tf_signal}
\end{figure}

Six different signal transformation recognition tasks are performed as pretext tasks \cite{multitask_selfsupervised, um2017data}. The details of the parameters associated with these tasks (for instance signal-to-noise ratio, scaling factor, stretching factor, and others) are tuned based on a large number of experiments and are explored in depth in Section \ref{Analysis of Individual Transformation Task} and \ref{Analysis of Multiple Transformation Task}. These transformations are described below and a sample of transformed signals with automatically generated labels is illustrated in Figure \ref{fig:tf_signal}.

\textit{Noise addition:} In this transformation, random noise from a Gaussian distribution, $N(t)$, is added to the original ECG signal $S(t)$. $N(t)$ is obtained from a random noise generator where the mean of the distribution is set to $0$, and the standard deviation to $\sqrt{E_{N_{avg}}}$. Here, the average power of $N(t)$, $E_{N_{avg}}$ is calculated as $10^{(E_{S_{avg}} - \alpha)/10}$, where $E_{S_{avg}}$ is the average power of $S(t)$ and $\alpha$ is the desired Signal to Noise Ratio (SNR). Finally the noise-added signal is generated as $S(t) + N(t)$.

\textit{Scaling:} The original magnitude of the ECG $S(t)$ is transformed as $\beta \times S(t)$, where $\beta > 0$ and $\beta$ is the manually assigned scaling factor.

\textit{Negation:} The original amplitude of the ECG $S(t)$ is multiplied by $-1$, causing a spatial inversion of the time-series. The transformed signal can be mathematically expressed as $-S(t)$. 

\textit{Temporal Inversion:} Given the original ECG signal $S(t)$, where $t = 1, 2, ..., N$ and $N$ is the length of the time-series, the temporally inverted version of the signal is expressed as $S'(t)$, where $t = N, N-1, ..., 1$.

\textit{Permutation:} In this transformation, the original ECG, $S(t)$, is divided into $m$ segments and shuffled, randomly perturbing the temporal location of each segment. Let's assume $S(t)$ is expressed as $S(t) = [s_i(t) | i = 1, 2, ..., m]$, a sequence of segments with segment numbers $i = 1, 2, ..., m$. Accordingly, the permuted signal $S_{p}(t)$ can be obtained as $S_{p}(t) = [s_{i}(t)]$, where the sequence of $i = 1, 2, ..., m$ is randomly shuffled.

\textit{Time-warping:} In this transformation, randomly selected segments of the original ECG signals $S(t)$ are stretched or squeezed along the $x$ (time) axis. Let's assume $\Phi(S(t), k)$ is an interpolation-based time-warping function where $k$ is the stretch factor (with the corresponding squeeze factor represented as $1/k$). If the signal, $S(t)$, is expressed as $S(t) = [s_i(t) | i = 1, 2, ..., m]$, a sequence of segments with segment numbers $i = 1, 2, ..., m$, the time-warped (transformed) signal $T(t)$ is obtained by applying $\Phi(s_{i}(t), k)$ to half of the segments selected randomly, and $\Phi(s_{j}(t), 1/k)$ to the other half of the segments where $i \neq j$. Where $T(t)$ becomes longer or shorter than $S(t)$ due to the even or odd nature of $m$, $T(t)$ is clipped or zero-padded accordingly to maintain the original input length. 

All of the above-mentioned transformations are applied to the original signals, the outcome of which are stacked to create the input matrix $X$, while the corresponding labels of the transformations $0, 1, ..., 6$ are stacked to create the corresponding output matrix $P$, with $0$ denoting the original matrix and integers $1$ to $6$ indicating the $6$ transformations. The input-output matrices are then shuffled to re-order the transformations and their corresponding outputs. 

The architecture of the proposed network for learning ECG representations using the created transformation dataset described above consists of $3$ convolutional blocks (conv-block) as \textit{shared layers}, and $7$ branches, each with $2$ dense layers, as \textit{task-specific layers}. Each conv-block is made of $2\times1$D convolution layers with ReLu activation functions followed by a max-pooling layer of size $8$. The number of filters are increased from $32$ to $64$ and $128$, whereas, the kernel size is decreased from $32$ to $16$ and $8$ respectively. Global max-pooling operation is performed after the final conv-block. Next, successive to flattening the embedding, the output is fed to $2$ fully connected layers with $128$ hidden nodes and ReLu activation functions. To overcome possible overfitting, $60\%$ dropout is introduced. Finally, the model output is generated through a sigmoid layer. The summary of the network architecture is presented in Table \ref{tab:SSL_arch}. Through learning the set of synthesized input-outputs described above, the network can learn strong ECG representations, which can be obtained by flattening the feature embeddings extracted after the shared convolutional blocks.

\subsection{Emotion Recognition}
In the second stage ($T_{d}$) of learning, the model is trained with the true emotion labels ${y}_{i}$ for emotion classification. $R_{ECG}$ obtained from $T_{p}$ contain useful information regarding the original signals $X_{0}$, which are used to perform $T_{d}$. To utilize $R_{ECG}$ towards performing $T_{d}$, a simple network $\rho = w(R_{ECG}, {y}_{i}, \theta')$ is employed, with $\rho$ as the probability vector of $T_{d}$ classes, and $\theta'$ as the set of trainable parameters. Finally, we calculate the optimum value of $\theta'$ by minimizing the cross entropy loss: 
\begin{equation}
    \begin{aligned}
        {L} = \sum_{i=1}^{M}{ {y}_{i} \log{\rho_i}},
    \end{aligned}
    \label{equ:ds_loss}
\end{equation}
where $M$ is total number of emotional classes. 

The emotion recognition network contains convolution layers similar to those used in learning ECG representation, followed by fully connected layers with $512$ hidden nodes. The weights of the shared layers of the signal transformation recognition network are frozen and transferred to the emotion recognition network. However, the fully connected layers of the emotion recognition network are not carried over from the transformation recognition network, and are instead trained using the labelled dataset (fully supervised) to perform emotion recognition. It should be noted that as we use different datasets in this study, the fully-connected layers are fine-tuned for each dataset. Specifically, $2$ dense layers are utilized for SWELL and WESAD datasets, while $3$ dense layers with L2 regularization ($0.0001$) and dropout of $40\%$ and $20\%$ are used for AMIGOS and DREAMER datasets respectively. The final output is taken from a sigmoid (binary classification) or softmax (multi-class classification) layer. We intentionally kept the fully-connected layers simple (single-task as opposed to multi-task) and somewhat shallow, in order to be able to evaluate the ability of the self-supervised approach to learn robust generalized ECG representations. The overview of the self-supervised solution is presented in Figure \ref{fig:architecture}, where successive to a self-supervised learning ECG representations, transfer learning \cite{7404017} is used for emotion recognition with the fully-supervised model.

\section{Experiments} \label{setup} 

\subsection{Datasets}
We used four publicly available datasets to evaluate the proposed solution in depth on a large number of different subjects, in different circumstances and under different data collection protocols, and using different hardware. The important characteristics of the datasets are summarized in Table \ref{tab:dataset_desc} and a brief description of each dataset is provided bellow.

\begin{table}[h]
    \centering
    \caption{The summary of the four datasets used are presented.}
    \begin{tabular}{l l l l}
     \hline 
         \textbf{Dataset} &  \textbf{Participants} & \textbf{Attributes} & \textbf{Classes} \\ \hline\hline
         AMIGOS & 40 & \makecell[l]{Arousal \\ Valence \\} & \makecell[l]{$9$  \\ $9$} \\ \hline
         DREAMER & 23 & \makecell[l]{Arousal \\ Valence } & \makecell[l]{$5$  \\ $5$ } \\ \hline
         WESAD & 17 & \makecell[l]{Affect State \\} & \makecell[l]{$4$ \\} \\ \hline
         SWELL & 25 & \makecell[l]{Stress \\Arousal \\ Valence \\}  & \makecell[l]{$3$  \\$9$  \\ $9$ \\}\\ \hline
    \end{tabular}
    \label{tab:dataset_desc}
\end{table}

\subsubsection{AMIGOS} 
The AMIGOS dataset \cite{amigos_dataset} was collected from $40$ participants to study personality, mood, and affective responses, while engaging with multimedia content in two different contexts: $a)$ alone, and $b)$ in a group of four people. To perform this study, participants were shown short and long emotional video clips to elicit affective states. The short video clips had a length of $250$ seconds, whereas the long video clips were longer than $14$ minutes. ECG data were recorded using Shimmer sensors \cite{shimmer} at a sampling frequency of $256$ Hz. A total of $16$ video clips were shown to each participant, and self-assessed arousal and valence scores were recorded after each trial on a scale of $1$ to $9$.

\subsubsection{DREAMER} 
The DREAMER dataset \cite{dreamer_dataset} was collected from $23$ participants while they were presented with audio and video stimuli in the form of movie clips in order to elicit emotional reactions. A total of $18$ video clips were shown, inducing $9$ different emotions namely amusement, excitement, happiness, calmness, anger, disgust, fear, sadness, and surprise. Each clip was $60$ seconds long. In addition, neutral clips were shown before each session to help subjects return to a neutral emotional state. ECG signals were collected using shimmer sensors \cite{shimmer} at a sampling rate of $256$ Hz. Subjective experience of arousal and valence scores were collected using self-assessment manikins \cite{bradley1994measuring}. At the end of each session, both arousal and valence scores were recorded for the entire session on a scale of $1$ to $5$, ranging from uninterested/bored to excited/alert for arousal and unpleasant/stressed to happy/elated for valence.

\subsubsection{WESAD} 
The dataset for WEarable Stress and Affect Detection (WESAD) contains ECG data from $17$ participants. A RespiBAN Professional \cite{respiban} sensors were used to collect ECG at a sampling rate of $700$ Hz. The goal was to study four different affective states namely neutral, stressed, amused, and meditated. To perform this study, four different test scenarios were created. First, $20$ minutes of neutral data were collected, during which participants were asked to do \textit{normal} activities such as reading a magazine and sitting/standing at a table. During the \textit{amusement} scenario, participants watched $11$ funny video clips for a total length of $392$ seconds. Next, participants went through public speaking and arithmetic tasks for a total of $10$ minutes as part of the \textit{stress} scenario. Finally, participants went through a guided \textit{meditation} session of $7$ minutes in duration. Upon completion of each trial, the ground truth labels for the affect states were collected using the Positive and Negative Affect Schedule (PANAS) scheme \cite{watson1988development}.

\subsubsection{SWELL} 
This dataset was created with the goal of understanding employees' mental stress or emotional attributes in a typical office environment under different conditions \cite{swell_dataset}. Three types of scenarios namely, normal, time pressure, and interruptions were designed. Under the \textit{normal} condition, participants were allowed to work on different tasks, for example preparing reports, making presentations, and so on, which were carried out over a maximum duration of $45$ minutes. During the \textit{time pressure} condition, the allowed time was reduced to $30$ minutes to induce pressure. In the \textit{interruption} session, participants were interrupted by sending emails and messages, with the participants being asked to respond to these messages. ECG signals were collected from $25$ participants using a TMSI MOBI device \cite{TMSi_Mobi} with self-adhesive electrodes at a sampling rate of $2048$ Hz. Finally, at the end of each scenario, the participants' self-reported affect scores were collected on a scale of $1$ to $9$.

\subsection{Data Pre-processing} \label{sub:data_preprocessing}
Since the above-mentioned datasets have been collected using different hardware, the signals have different spatiotemporal properties such as spatial range and sampling rate. To minimize the effects of such inter-dataset variations and discrepancies, three simple pre-processing steps were taken. First, SWELL and WESAD ECG signals are downsampled to $256$ Hz to be consistent with AMIGOS and DREAMER. Next, we remove baseline wander from all the four datasets by applying a high-pass IIR filter with a pass-band frequency of $0.8$ Hz. Lastly, we perform user-specific \textit{z}-score normalization. While a number of other pre-processing operations such as feature extraction could have been done, we intentionally kept the pre-processing minimal and simple in order to better understand the impact of the proposed model on learning important ECG representations based on almost raw input. Finally, the filtered ECG signals are segmented into a fixed window size of $10$ seconds and stacked into an array. No overlap is designated between segments to avoid any potential data leakage between training and testing data. It should be noted that the selection of the window size was empirical. Prior works utilizing these datasets have selected a wide range of different window sizes. For example, \cite{amigos_dataset} and \cite{santamaria2018using} selected a window length of $20$ seconds for AMIGOS, whereas \cite{dreamer_dataset} and \cite{sriramprakash2017stress} selected a window of $60$ seconds for both DREAMER and SWELL. As other examples, \cite{wesad} has selected $5$-second windows for WESAD, while \cite{lin2019explainable} has selected $1$-second windows for the same dataset.

\subsection{Implementation and Training}
In order to train the model, successive to pre-processing, each segment is used to generate the $6$ transformations described earlier. Finally, the original ECG signals along with the transformed signals are used to train the signal transformation recognition network. We implement the proposed architecture using TensorFlow. We share the implementation of the self-supervised network\footnote{\url{https://code.engineering.queensu.ca/17ps21/SSL-ECG}}.

\begin{figure}[t]
    \centering
    \includegraphics[width=1\columnwidth]{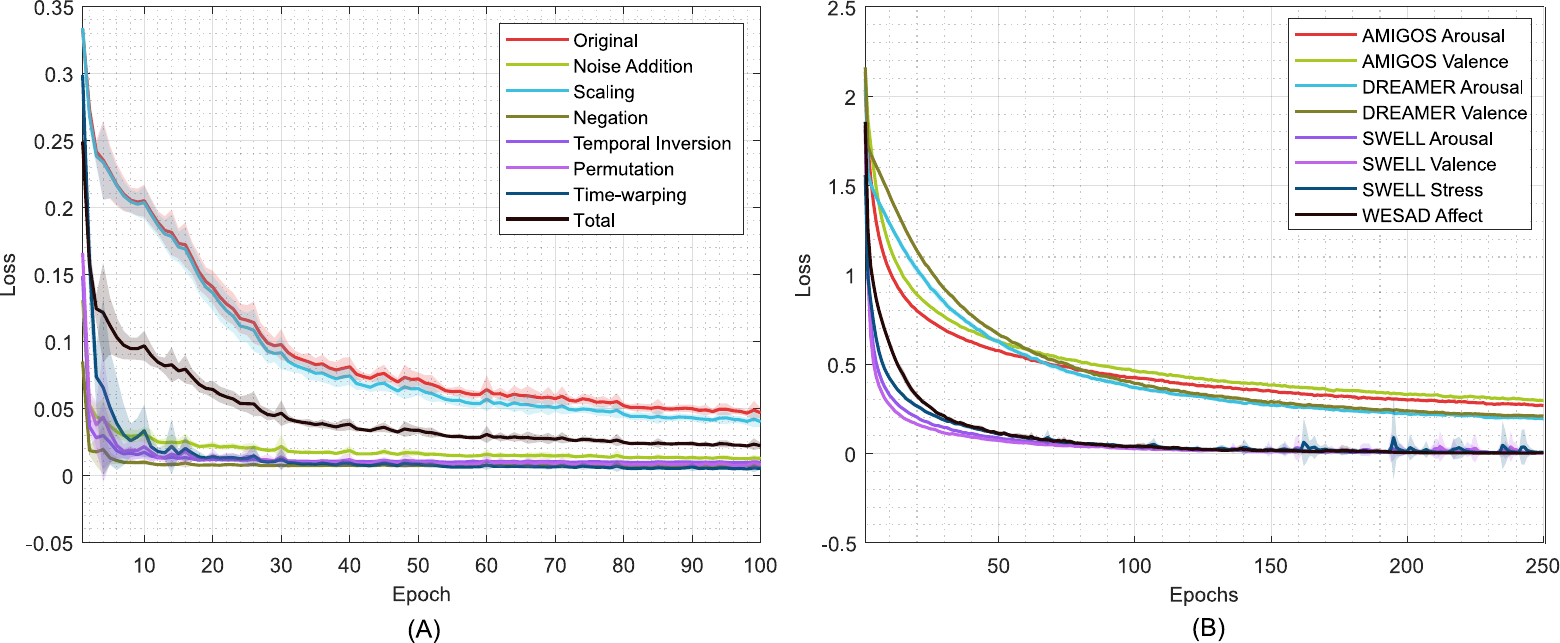}
    \caption[c]{Training losses vs. training epochs are presented for both networks, the of signal transformation network (left) and the emotion recognition network (right). In (A), the average and standard deviation of individual losses and the total output loss of the signal transformation recognition network are obtained from 10-fold training, while in (B), the average and standard deviation of individual losses for the different emotion recognition tasks obtained with 10-fold training for each dataset are presented.}
    \label{fig:all_loss}
\end{figure}

Similar to other works in this area \cite{amigos_dataset, santamaria2018using, dreamer_dataset, wesad, lin2019explainable, sriramprakash2017stress}, we use $10$-fold cross-validation to evaluate the performance of the model successive to shuffling of the pre-processed dataset. We randomly select 90\% of the data for training and the remaining 10\% is used for testing (this process is repeated 10 times without repeating the shuffling step). To train both the signal transformation recognition network and the emotion recognition network, Adam optimizer \cite{kingma2014adam} is used with a learning rate of $0.001$ and batch size of $128$. The signal transformation recognition network is trained for $100$ epochs, while the emotion recognition network is trained for $250$ epochs. The number of training epochs for each network is selected to enable the training reach a steady state. Figure \ref{fig:all_loss}(A) and (B) show the loss vs. training epoch for the transformation recognition and emotion recognition networks respectively during training. Figure \ref{fig:all_loss}(A) shows that the loss for temporal inversion, negation, permutation, and time-warping reach steady states earlier than the other transformations. Therefore, the network is trained until all the individual losses reach their respective steady states. Figure \ref{fig:all_loss}(B) shows that the different datasets reach their training steady states at different epochs. For example, the loss of WESAD and SWELL become stable well before $100$ epochs, while the steady states of AMIGOS and DREAMER are achieved after $200$ epochs. In summary, Figure \ref{fig:all_loss} shows that both pretext tasks and emotion recognition tasks train well and reach steady states using the proposed solution.

\section{Results} \label{eval}

\begin{table}[t]
    \centering
    \caption{The average and standard deviations of the accuracies and F1 scores for signal transformation recognition across the four datasets are presented.}
    \begin{tabular}{l|c|c}
    \hline
    \textbf{\multirow{2}{*}{Transformation}} & \multicolumn{2}{c}{\textbf{All datasets combined}} \\ \cline{2-3}
        & Acc. & F1  \\ \hline \hline
        
        \multirow{1}{*}{\textbf{Original}}  & $0.980\pm0.003$ & $0.927\pm0.007 $\\
        
        \multirow{1}{*}{\textbf{Noise Addition}} & $0.995\pm0.000$ & $0.979\pm0.003 $ \\
        
        \multirow{1}{*}{\textbf{Scaling}} & $0.982\pm0.003$ & $0.932\pm0.010 $ \\
        
        \multirow{1}{*}{\textbf{Temporal Inversion}} & $0.998\pm0.000$ & $0.992\pm0.004 $ \\

        \multirow{1}{*}{\textbf{Negation}} & $0.998\pm0.000$  & $0.990\pm0.000$ \\
        
        \multirow{1}{*}{\textbf{Permutation}} & $0.998\pm0.000$  & $0.989\pm0.003$ \\
        
        \multirow{1}{*}{\textbf{Time-warping}} & $0.997\pm0.003$  & $0.992\pm0.006$ \\ \Xhline{1pt}
        
        \multirow{1}{*}{\textbf{Average}} & $0.992\pm0.001$  & $0.972\pm0.005$ \\ \hline
    \end{tabular}
    \label{tab:ssl_result_signal_combined}
\end{table}

\begin{table*}[t]
    \centering
    \scriptsize
    \setlength\tabcolsep{4.5pt}
    \caption{Results for the individual transformation recognition tasks in each of the four datasets are presented.}
    \begin{tabular}{l|c|c|c|c|c|c|c|c}
    \hline
    \textbf{\multirow{2}{*}{Transformation}} & \multicolumn{2}{c|}{\textbf{AMIGOS}} & \multicolumn{2}{c|}{\textbf{DREAMER}} & \multicolumn{2}{c|}{\textbf{WESAD}} & \multicolumn{2}{c}{\textbf{SWELL}}  \\ \cline{2-9}
        & Acc. &  F1  & Acc. &  F1 & Acc. &  F1 & Acc. &  F1 \\ \hline \hline
        \multirow{1}{*}{\textbf{Original}}                   & $0.973\pm0.003$ & $0.901\pm0.010 $ & $0.986\pm0.005$ & $0.953\pm0.020 $ & $0.981\pm0.008$ & $0.950\pm0.014 $    & $0.984\pm0.003$ & $0.941\pm0.009 $   \\ 
        
        \multirow{1}{*}{\textbf{\makecell{Noise Addition}}}  & $0.987\pm0.002$ & $0.954\pm0.008 $ & $1.000\pm0.000 $ & $1.000\pm0.000 $ & $1.000\pm0.000 $    & $1.000\pm0.000 $ & $0.999\pm0.000 $   & $0.999\pm0.000$ \\ 
        
        \multirow{1}{*}{\textbf{Scaling}}                    & $0.975\pm0.003$ & $0.904\pm0.020 $ & $0.986\pm0.005$ & $0.955\pm0.024 $ & $0.984\pm0.008$ &  $0.952\pm0.012 $     & $0.987\pm0.001$  & $0.948\pm0.006 $   \\ 
        
        \multirow{1}{*}{\textbf{\makecell{Temporal Inversion}}}         & $0.997\pm0.001$ & $0.988\pm0.004 $  & $1.000\pm0.000 $ & $1.000\pm0.000 $   & $0.999\pm0.001$ & $1.000\pm0.000 $   & $1.000\pm0.000 $   & $1.000\pm0.000 $\\ 

        \multirow{1}{*}{\textbf{Negation}}                   & $0.996\pm0.001$ & $0.987\pm0.005 $ & $1.000\pm0.000 $ & $1.000\pm0.000 $ & $0.999\pm0.002$ &  $0.995\pm0.007 $  & $1.000\pm0.000 $  & $0.998\pm0.004 $   \\ 
        
        \multirow{1}{*}{\textbf{Permutation}}                & $0.996\pm0.001$  & $0.984\pm0.005 $ & $0.999\pm0.000$ & $0.999\pm0.003 $ & $0.999\pm0.001$ & $0.994\pm0.007 $    & $0.998\pm0.001$ & $0.991\pm0.009 $   \\ 
        
        \multirow{1}{*}{\textbf{Time-warping}}               & $0.998\pm0.001$ & $0.993\pm0.007 $ & $0.998\pm0.003 $  & $0.999\pm0.003$ & $0.993\pm0.009$ & $0.991\pm0.006 $    & $0.996\pm0.004$ & $0.990\pm0.005 $   \\\hline 
        
        \multirow{1}{*}{\textbf{Average}}                    & $0.989\pm0.002$ & $0.959\pm0.008 $ & $0.996\pm0.002$ & $0.987\pm0.004 $ & $0.994\pm0.004$ & $0.983\pm0.007 $     & $0.995\pm0.001$ & $0.981\pm0.005 $  \\ \hline
    \end{tabular}
    \label{tab:ssl_result_signal_individual}
\end{table*}

\subsection{Signal Transformation Recognition}
Table \ref{tab:ssl_result_signal_combined} and \ref{tab:ssl_result_signal_individual} presents the F1 scores and accuracies of our proposed network for transformation recognition. Results shown in Table \ref{tab:ssl_result_signal_combined} are calculated on all the datasets combined. The results show very high F1 scores and accuracies for all the transformations as well as the original signal. F1 scores $0.927$ and $0.932$ are achieved when detecting the original signal and the scaling transformations. Next, relatively higher F1 score of $0.979$ is achieved for recognition of noisy signals. We notice few tasks consistently report very high scores ($\approx 0.99$), namely temporal inversion, negation, permutation, and time-warping. For accuracy, very high values (greater than $0.98$) are achieved for every transformation. Beside high F1 scores and accuracies, we also achieve very low standard deviations for all the transformation recognition tasks, which indicates a consistent performance by the network in all the training folds. Finally, an average F1 score of $0.972$ and an average accuracy of $0.992$ are reported for all the tasks combined.

Next, we further investigate the consistency and generalization of the performance of the transformation recognition network across different datasets. Table \ref{tab:ssl_result_signal_individual} presents the results on each of the four datasets, showing very high F1 scores and accuracies. This is a strong indicator of the generalizability of our approach as the model performs consistently on all the datasets with F1 scores of $0.959$ to $0.987$ and accuracies of $0.989$ to $0.996$, with low standard deviations in every case.

\begin{table}
    \centering
    \caption{The emotion categories, number of classes for each class, and multi-class emotion recognition results are presented for each of the four datasets.}
    \begin{tabular}{l|l|l|c|c}  
    \hline    \textbf{Dataset} & \textbf{Attribute} & \textbf{\makecell{Classes}} & \textbf{Acc.} & \textbf{F1} \\ \hline \hline
        
        \multirow{2}{*}{\textbf{AMIGOS}} & Arousal & $9$ & $0.796$ & $0.777$ \\ \cline{2-5}
        & Valence & $9$ & $0.783$ & $0.765$ \\ \hline
        
        \multirow{2}{*}{\textbf{DREAMER}} & Arousal & $5$ & $0.771$ & $0.740$ \\ \cline{2-5}
        & Valence &  $5$ &  $0.749$ & $0.747$ \\ \hline
        
        \multirow{1}{*}{\textbf{WESAD}} & Affect State & $4$ & $0.950$ & $0.940$ \\ \hline

        \multirow{3}{*}{\textbf{SWELL}} & Arousal &  $9$ &  $0.926$ & $0.930$ \\ \cline{2-5}
        & Valence &  $9$ &  $0.938$ & $0.943$ \\ \cline{2-5}
        & Stress &  $3$ & $0.902$ & $0.900$ \\ \hline
        
    \end{tabular}
    \label{tab:ssl_result_emoti}
\end{table}

\subsection{Emotion Recognition}
Results obtained from the emotion recognition network are presented in Table \ref{tab:ssl_result_emoti}. Arousal and valence detection are performed with three datasets AMIGOS, DREAMER, and SWELL, stress detection is performed with SWELL, and classification of different affective states are performed in WESAD, as per the availability of output labels. As presented in Table \ref{tab:dataset_desc}, the four datasets contain different number of classes for each target output. Therefore, the classification tasks are performed accordingly, the accuracies and F1 scores of which are presented in Table \ref{tab:ssl_result_emoti}. It should be noted that to the best of our knowledge, no previous work has utilized AMIGOS, DREAMER and SWELL datasets for multi-class classification. this is the first time, multi-class classification of arousal and valence score is attempted on. With AMIGOS, we achieve accuracies of $79.6\%$ and $78.3\%$ for arousal and valence respectively, where $9$-class classification is performed. In DREAMER, $5$-class classification is performed, achieving accuracies of $77.1\%$ and $74.9\%$ for arousal and valence respectively. Next, we perform detection of $4$ affective states using WESAD, reporting an accuracy of $95.0\%$. Lastly, arousal, valence, and stress detection is performed on SWELL, achieving accuracies of $92.6\%$, $93.8\%$, and $90.2\%$ respectively, where arousal and valence scores have $9$ classes while stress has $3$ classes.

\noindent \textbf{Comparison:}
To further evaluate the proposed framework, we compare our model's performance to past work as shown in Table \ref{tab:comparison_table}. It should be noted that most of the prior works using the four studied datasets have performed two-class or three-class classifications \cite{amigos_dataset, dreamer_dataset, swell_dataset, wesad} instead of utilizing all the multiple classes available for each dataset. Therefore, in order to perform a fair comparison, in addition to the multi-class classifications presented in Table \ref{tab:ssl_result_emoti}, in Table \ref{tab:comparison_table} we present the results of the model applied to two/three-class classifications in accordance to the prior works and state-of-the-art methods presented in this table. In Table \ref{tab:comparison_table}, AMIGOS, DREAMER, and SWELL have all been classified in two-class format by setting a threshold at the mean output value dividing the outputs to high and low, while WESAD has been classified in three classes by ignoring the meditated output class. It should be noted that the works mentioned in Table \ref{tab:comparison_table} have also adopted a $10$-fold cross-validation scheme similar to the validation scheme used in this paper. Moreover, for each dataset, we have also carried out \textit{fully-supervised} classification by skipping the self-supervised step and directly training the emotion recognition network using the input ECG and emotion labels. In \cite{santamaria2018using}, a CNN was implemented to perform emotion classification on AMIGOS. It reported accuracies of $81\%$ and $71\%$ for classification of arousal and valence respectively. The results in Table \ref{tab:comparison_table}(A) show that the self-supervised CNN achieves accuracies of $88.9\%$ and $87.5\%$ for classification of arousal and valence respectively, outperforming the past works as well as the fully-supervised baseline CNN. In \cite{dreamer_dataset}, classification of arousal and valence was performed with DREAMER. An SVM classifier was implemented, reporting an accuracy of $62.4\%$ for both arousal and valence. Table \ref{tab:comparison_table}(B) shows that the proposed solution achieves considerable improvement over the state-of-the-art and the baseline CNN with accuracies of $85.9\%$ and $85\%$ for arousal and valence respectively. In a study on WESAD, \cite{lin2019explainable} proposed a multimodal CNN that was able to achieve state-of-the-art with an accuracy of $83\%$ while using only ECG. As shown in the Table \ref{tab:comparison_table}(C), our proposed model is able to outperform the CNN in \cite{lin2019explainable} as well as the baseline CNN with accuracy of $96.9\%$. Lastly, a study on SWELL performed stress detection \cite{sriramprakash2017stress}, in which two classifiers, \textit{k}NN and SVM, were utilized. The results in Table \ref{tab:comparison_table}(D) show that the baseline CNN slightly outperforms the SVM, whereas the self-supervised model considerably outperforms both with an accuracy of $93.3\%$. Additionally, we perform two-class classification of arousal and valence on SWELL for the first time, achieving very accurate results.

It should be noted that in order to perform a fair comparison, Table \ref{tab:comparison_table} excludes prior works such as: \textit{i}) \cite{harper2019bayesian} and \cite{siddharth2019utilizing} which have used different validation schemes; \textit{ii}) \cite{liu2019multimodal, koldijk2016detecting, verma2019comprehensive} which have used multiple modalities (ECG with EEG or GSR) available in some of the datasets; and \textit{iii}) \cite{alberdi2018using}, which has formulated the problem as regression instead of classification.

\renewcommand{\tabcolsep}{5pt}

\begin{table*}[t]
\centering
\caption{The results of the self-supervised model on all the datasets are presented and compared with prior works including the state-of-the-art, as well as a fully-supervised CNN as a baseline.}
\begin{minipage}[b]{0.45\textwidth}
    \centering
    \caption*{\textbf{A: AMIGOS}}
    \begin{tabular}[t]{l|l|l|l|l|l}
        \hline 
        \multirow{2}{*}{\textbf{Ref.}} & \multirow{2}{*}{\textbf{Method}} & \multicolumn{2}{l|}{\textbf{Arousal}} & \multicolumn{2}{l}{\textbf{Valence}} \\ \cline{3-6}
             & & \textbf{Acc.} & \textbf{F1} & \textbf{Acc.} & \textbf{F1} \\ \hline\hline
            \cite{amigos_dataset} & GNB & --
            & $0.545$ & --
            & $0.551$ \\ \hline
            \cite{santamaria2018using} & CNN & $0.81$  & $0.76$ & $0.71$  & $0.68$ \\ \hline 

            \multirow{2}{*}{\textbf{Ours}} & \makecell[l]{Fully-Supervised CNN} & $0.844$  & $ 0.835$ & $0.811$  & $0.809$ \\ \cline{2-6}
             & \textbf{\makecell[l]{Self-Supervised CNN}} & $\textbf{0.889}$  & $\textbf{0.884}$ & $\textbf{0.875}$  & $\textbf{0.874}$ \\ \hline
    \end{tabular}
\end{minipage}
\hfillx
\begin{minipage}[b]{0.45\textwidth}
    \centering
    \caption*{\textbf{B: DREAMER}}
    \begin{tabular}[t]{l|l|l|l|l|l}
        \hline
            \multirow{2}{*}{\textbf{Ref.}} & \multirow{2}{*}{\textbf{Method}} & \multicolumn{2}{l|}{\textbf{Arousal}} & \multicolumn{2}{l}{\textbf{Valence}} \\ \cline{3-6}
             && \textbf{Acc.} & \textbf{F1} & \textbf{Acc.} & \textbf{F1} \\ \hline\hline
            \cite{dreamer_dataset} & SVM & $0.624$  & $0.580$ & $0.624$  & $0.531$ \\ \hline
            \multirow{2}{*}{\textbf{Ours}} & \makecell[l]{Fully-Supervised CNN} & $0.707$  & $ 0.708$ & $0.666$  & $0.658$ \\ \cline{2-6}
             & \textbf{\makecell[l]{Self-Supervised CNN}} & $\textbf{0.859}$  & $\textbf{0.859}$ & $\textbf{0.850}$  & $\textbf{0.845}$ \\ \hline %
    \end{tabular}
\end{minipage}    
\vfillx
\begin{minipage}[b]{0.35\textwidth}
    \centering
    \caption*{\textbf{C: WESAD}}
    \begin{tabular}[t]{l|l|l|l}
        \hline
            \multirow{2}{*}{\textbf{Ref.}} & \multirow{2}{*}{\textbf{Method}} & \multicolumn{2}{l}{\textbf{Affect State}} \\ \cline{3-4}
             && \textbf{Acc.} & \textbf{F1} \\ \hline\hline
            \multirow{4}{*}{\cite{wesad}} & \textit{k}NN & $0.548$ & $0.478$ \\ \cline{2-4}
            & DT & $0.578$ & $0.517$ \\ \cline{2-4}
            & RF & $0.604$ & $0.522$ \\ \cline{2-4}
            & AB & $0.617$ & $0.525$ \\ \cline{2-4}
            & LDA & $0.663$  & $0.560$ \\ \hline
            \cite{lin2019explainable} & CNN & $0.83$  & $0.81$  \\ \hline 
            \multirow{2}{*}{\textbf{Ours}} & \makecell[l]{Fully-Supervised CNN} & $0.932 $  & $ 0.912 $ \\ \cline{2-4}
             & \makecell[l]{\textbf{Self-Supervised CNN}} & $\textbf{0.969}$  & $\textbf{0.963}$  \\ \hline
    \end{tabular}
\end{minipage}
\hfillx
\begin{minipage}[b]{0.55\textwidth}
    \centering
    \caption*{\textbf{D: SWELL}}
    \begin{tabular}[t]{l|l|l|l|l|l|l|l}
        \hline
            \multirow{2}{*}{\textbf{Ref.}} & \multirow{2}{*}{\textbf{Method}} & \multicolumn{2}{l|}{\textbf{Stress}} & \multicolumn{2}{l|}{\textbf{Arousal}} & \multicolumn{2}{l}{\textbf{Valence}} \\ \cline{3-8}
             & &\textbf{Acc.} & \textbf{F1} & \textbf{Acc.} & \textbf{F1} & \textbf{Acc.} & \textbf{F1} \\ \hline\hline
            \multirow{2}{*}{\cite{sriramprakash2017stress}} & \textit{k}NN & $0.769$ & -- & -- & -- & -- & -- \\ \cline{2-8}
            & SVM & $0.864$ & -- & -- & -- & -- & -- \\ \hline 
            \multirow{2}{*}{\textbf{Ours}} & \makecell[l]{Fully-Supervised CNN} & $0.894$ & $0.874$ & $0.956$ &  $0.962$ & $0.961$ & $0.956$ \\ \cline{2-8}  
             & \textbf{\makecell[l]{Self-Supervised CNN}} & $\textbf{0.933}$ & $\textbf{0.924}$ & $\textbf{0.967}$ & $\textbf{0.964}$ & $\textbf{0.973}$ & $\textbf{0.969}$ \\ \hline 
    \end{tabular}
\end{minipage}   
\label{tab:comparison_table}
\end{table*}

\section{Analysis and Discussion} \label{Discussion}

\begin{figure*}
    \centering
    \includegraphics[width=1\textwidth]{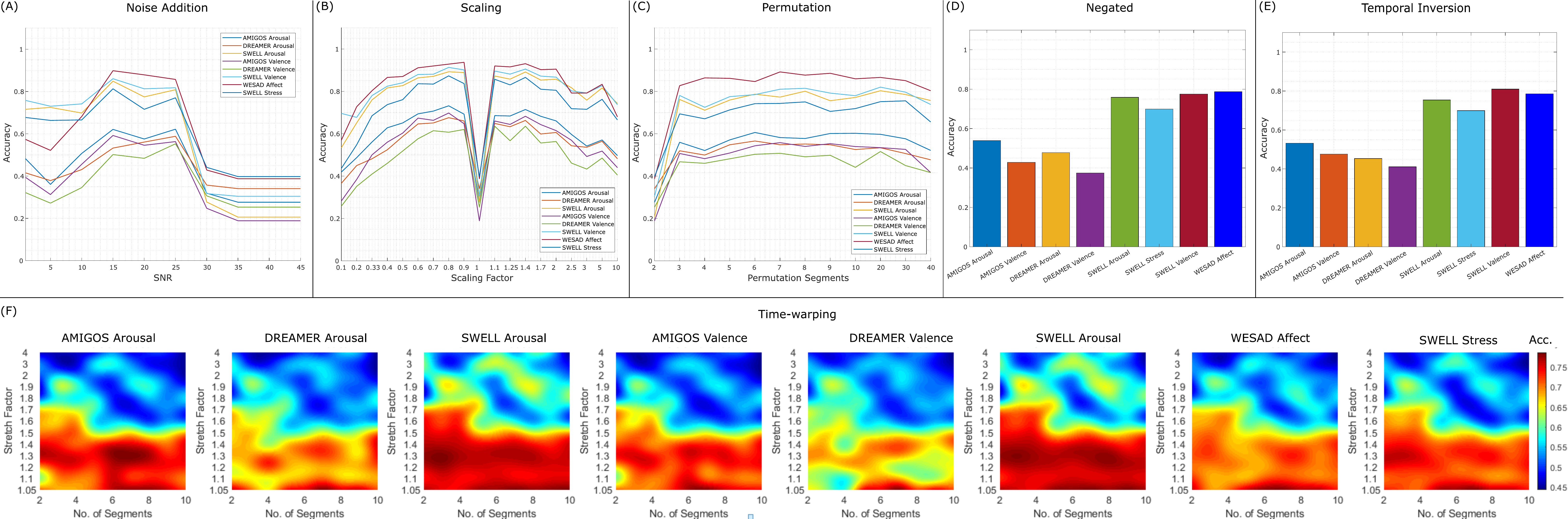}
    \caption{The impact of transformation parameters on emotion recognition accuracy is illustrated. Parts (A), (B), and (C) are $2$D graphs as noise addition, scaling, and permutation transformations only have one parameter. Parts (D) and (E) are presented in bar-plot format as temporal inversion and negation transformations are operations with no controlling parameter. Lastly, part (F) shows $3$D plots for time-warping as this operation has two parameters, with \textit{x} and \textit{y} axes representing the number of segments and stretch factor respectively, and the accuracy is denoted by the color intensity.}
    \label{fig:grid_sep}
\end{figure*}

\subsection{Single-task Self-supervision} \label{Analysis of Individual Transformation Task}
We study the impact of individual transformations on the emotion recognition performance. For this study, we create a single-task CNN, instead of a multi-task CNN. For this study, we select a range of possible values for the parameters that control the significance of each individual transformation. For noise addition, the noise amplitude parameter is changed to result in signal-to-noise ratios (SNR) of $2$ to $45$. For the scaling transformation parameter, scaling factors of $0.1$ to $10$ are selected. For permutation, the number of segments is varied between $2$ to $40$. For time-warping, the stretch factor is varied between $1.05$ to $4$, while the number of segments varies from $2$ to $40$. Finally, temporal inversion and negation, which do not have any controllable parameters are also studied. We should point out that these ranges of parameters are selected to create a wide range of transformed signals, almost identical to the original signal in one end and considerably different in the other. Figure \ref{fig:grid_sep} shows the emotion recognition performance for each dataset vs. the controllable parameters for each individual transformation task. To simplify the search, a single $90$-$10$ split for training-testing was performed instead of $10$-fold. This analysis provides in-depth insight into the effect of the parameters associated with each self-supervised recognition task (transformation) on the emotion recognition outcome. Furthermore, this analysis helps us narrow down the set of suitable transformation parameters in order to achieve the best performance. From Figure \ref{fig:grid_sep}(A), we notice that for SNR values of $15$ and $20$ the model shows highest performance for emotion recognition compared to SNR values greater than $30$ or less than $10$. Moreover, from Figure \ref{fig:grid_sep}(B), we notice that for the scaling factor of $1$, where the scaled signal and the original signals are identical, the model performance on emotion recognition is poor. However, for slightly greater or smaller scaling values, the model shows significantly better performance. Interestingly, the performance drops when the scaling factor is greater than $1.4$ or less than $0.6$. Figure \ref{fig:grid_sep}(C) illustrates that when utilizing permutations, poor performance is achieved when the input is divided into only $2$ segments or more than $20$ segments. Otherwise, when the number of segments varies in the range of $3$ to $20$, a relatively steady performance is observed. Time-warping analysis is presented in a $3$D plot in Figure \ref{fig:grid_sep}(F) where dark red regions show higher accuracies. It can be seen that number of segments in the range of $6$ to $9$ and a stretch factor in the range of $1.05$ and $1.35$ consistently result in better performance compared to other combinations.

In summary, the analysis above shows that for all the different transformation tasks, when the transformed signal is highly distorted compared to the original input, for example with high amplitude noise (SNR $< 10$), large number of segments ($> 20$), very high or very low scaling factor ($< 1.5$ or $> 0.6$), or large stretch factor ($> 1.4$), emotion recognition accuracy drops. This is due to the fact that distinguishing between the original signal and the highly distorted versions of the signal becomes simple. This results in the network not learning useful ECG representations. Similarly, when the transformed signal remains almost similar to the original signal, for instance when the SNR $> 35$, or a scaling factor in the range of $0.95$ to $1.05$, or the number of permutated segments equal to $2$, the performance also drops. This is due to the fact that the recognition tasks in such cases become too difficult for the network to properly learn the required ECG representations from. Hence, we conclude that there is a range of values for the parameters associated with the self-supervised tasks, for which these tasks are located in an suitable difficulty range, resulting in optimum learning.

\begin{figure*}
    \centering
    \includegraphics[width=1\textwidth]{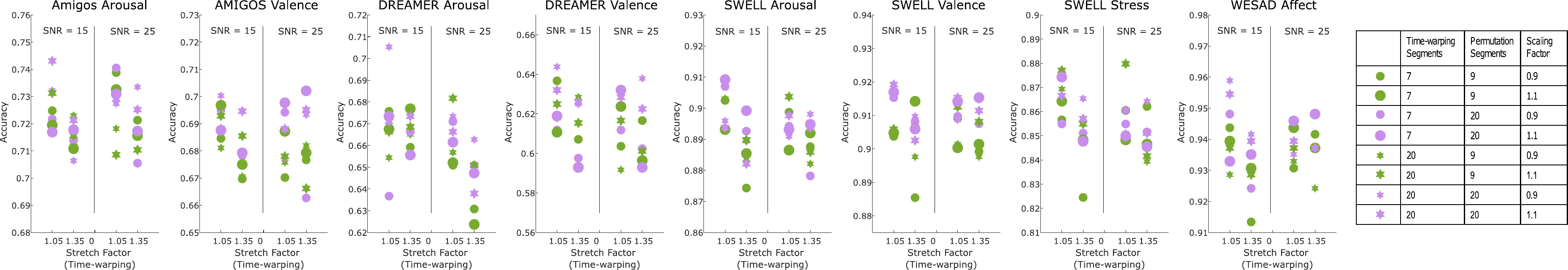}
    \caption{Emotion recognition results vs. $5$D vector of parameters controlling all $6$ tasks simultaneously are presented. The resulting $6$ dimensional plots are represented as follows: Y axis presents the emotion recognition accuracy, the two plots separated by the vertical line correspond to SNR, the time-warping stretch factor is represented by the X axis, the scaling factor, permutation segments, and time-warping segments are represented by marker size, color, and shape respectively.}
    \label{fig:grid_mt}
\end{figure*}

\begin{figure}
    \centering
    \includegraphics[width=1\columnwidth]{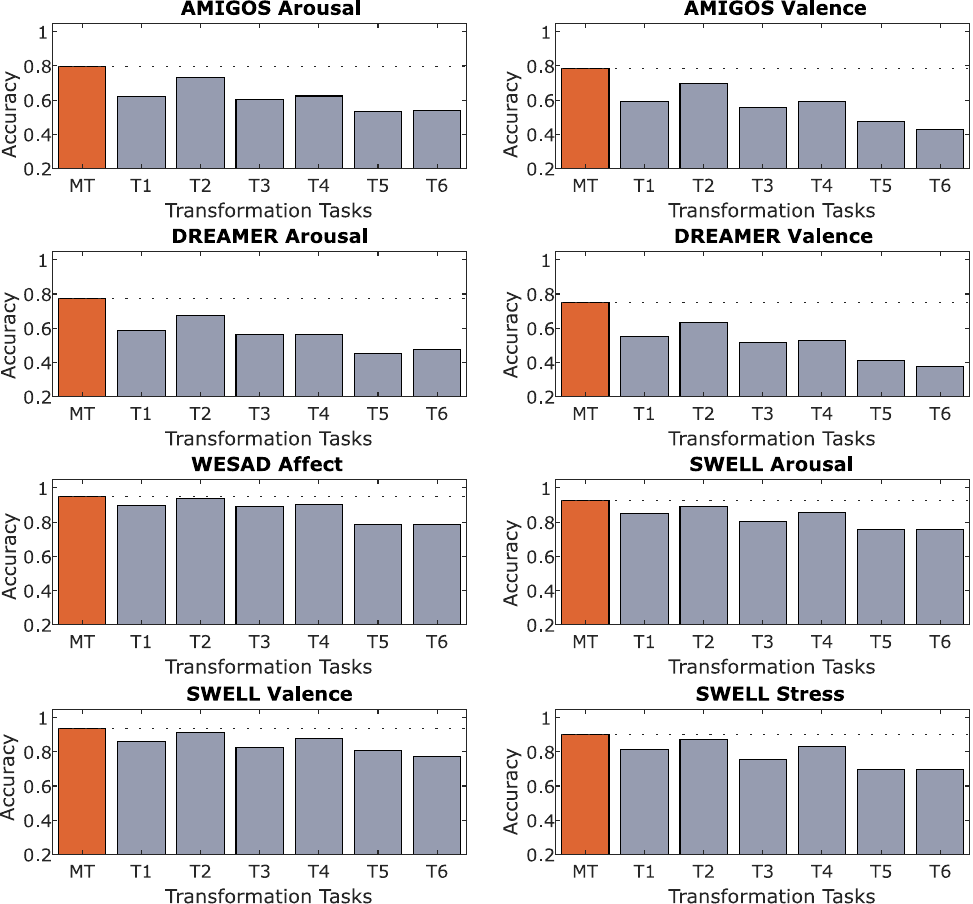}
    \caption{Comparison between single-task and multi-task self-supervision is presented. MT denotes multi-task, while T1 through T6 denote noise addition, scaling, permutation, time-warping, temporal inversion, and negation respectively.}
    \label{fig:STMT}
\end{figure}

\begin{figure}[t]
    \centering
    \includegraphics[width=1\columnwidth]{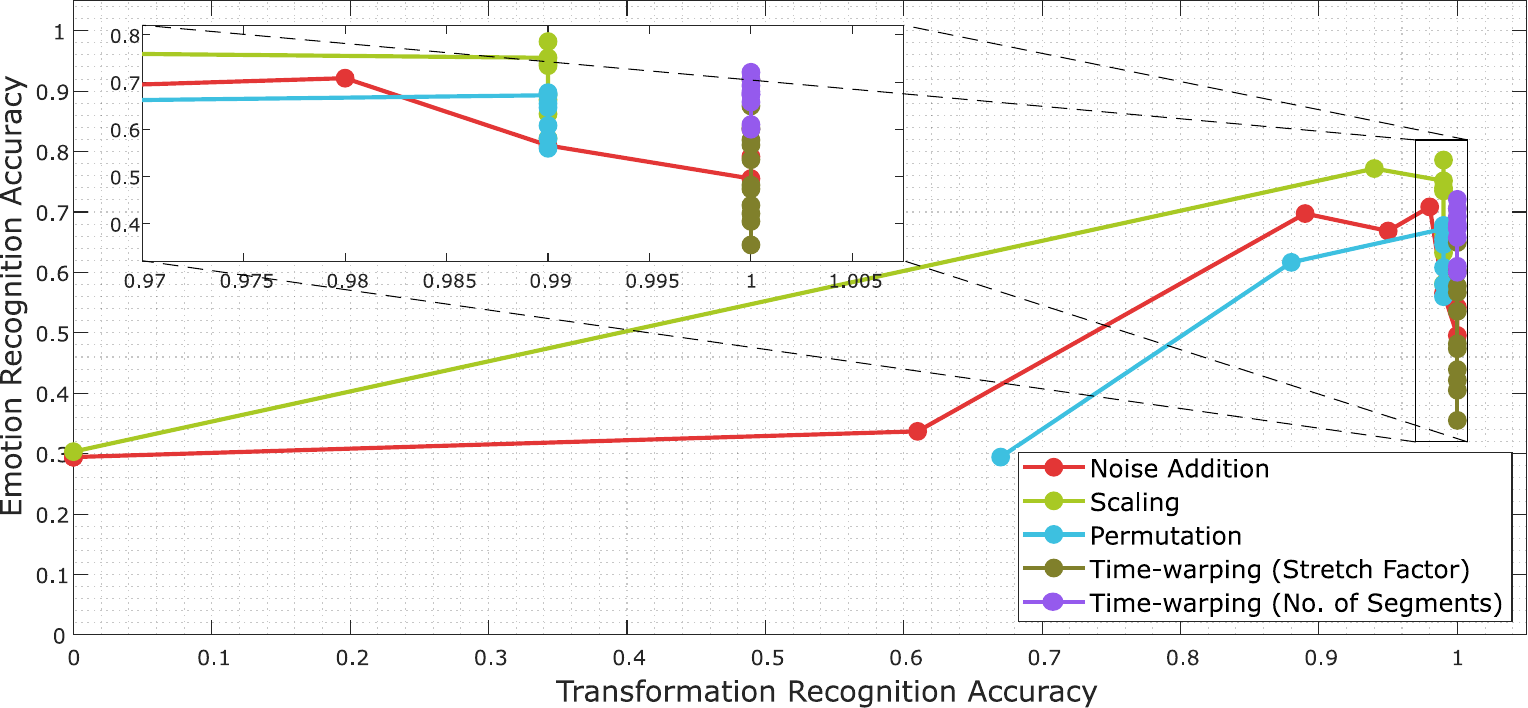}
    \caption{The relationship between emotion recognition accuracy and transformation recognition is presented. Time-warping (Stretch Factor) denotes that the stretch factor is varied while the number of segments is set to the optimum value. Similarly, Time-warping (No. of Segments) indicates that the number of segments is varied while the stretch factor is set to the optimum value.}
    \label{fig:ervsstr}
\end{figure}

\subsection{Multi-task Self-supervision} \label{Analysis of Multiple Transformation Task}
This section discusses the emotion recognition performance when multiple signal transformations (multi-task) are performed to learn ECG representations. The reason we believe this analysis is necessary despite already analyzing individual transformations in the previous subsection is that the use of a multi-task network may have an impact on the overall performance which might be different than the aggregation of several individual self-supervised tasks.
To perform this study, we use the results from the previous section (Subsection \ref{Analysis of Individual Transformation Task}) to narrow down the search given the large degree of freedom in the multi-task parameter space. Also, to further simplify the search, similar to Figure \ref{fig:grid_sep}, a single $90$-$10$ split for training-testing was performed instead of a complete $10$-fold. The results obtained from different combination of signal transformation parameters are presented in a six-dimensional plot as shown in Figure \ref{fig:grid_mt}. Our analysis shows that ($15, 0.9, 20, 9, 1.05$) is the optimal vector of parameters for SNR, scaling factor, number of permutation segments, number of time-warping segments, and time-warping stretch factor respectively. Our analysis further confirms that the results obtained from single-task self-supervision are in compliance with results obtained from multi-task self-supervision. For example, Figure \ref{fig:grid_sep}(A) shows that an SNR $= 15$ results in better emotion recognition compared to SNR $= 25$. Very similar results are also observed in Figure \ref{fig:grid_mt}, as SNR $= 15$ results in the best accuracy in $6$ out of $8$ instances. As another example, in both single-task and multi-task settings, a stretch factor of $1.05$ shows superior performance compared to $1.35$.

Figure \ref{fig:STMT} compares the emotion recognition results when ECG representations are learned using self-supervised single-tasks to when multi-task representation learning is used. These results are reported for the optimum parameters for each case. 
The figure shows that different transformation recognition tasks help the self-supervised model learn better high-level representations of ECG. As the multi-task network consistently outperforms single-task setups (more prominently in AMIGOS and DREAMER), it can be concluded that different transformations help the model in learning different aspects of ECG representations. For instance while noise addition, scaling, and negation may be contributing more towards learning spatial representations of ECG, others such as temporal inversion, permutation, and time-warping may be contributing more to learning the temporal relations. Moreover, with respect to the relationship between different transformation recognition and emotion recognition tasks, specific commonalities are observed. In particular, we see that scaling (T2) has greater impact on emotion recognition compared to noise addition (T1) and negation (T6). On the other hand, time-warping (T4) has greater impact compared to permutation (T3) and temporal inversion (T5).

In the multi-task network, the total loss is calculated as a weighted average of individual losses (see Eq. \ref{equ:total_loss}). We realize that these weights (loss coefficients) play an important role in learning the ECG representations since during training of the self-supervised network, different losses reach their steady states at different epochs. Specifically, the losses of the temporal inversion and spatial negation saturate very quickly and report almost perfect $F1$ scores just after $5$ to $7$ epochs. The possible reason for this phenomena is the clear difference between inverted and negated signals with respect to the original signal, making it very easy for the model to correctly classify these transformations. As a result, we utilize smaller weights for temporal inversion and negation losses compared to the other transformations. All the weights were set empirically with the goal of maximizing performance. The loss coefficients of $0.0125$ are used for temporal inversion and negation, while for the rest of the $5$ losses, the weights are set to $0.195$.

\begin{figure*}
    \centering
    \includegraphics[width=0.8\textwidth]{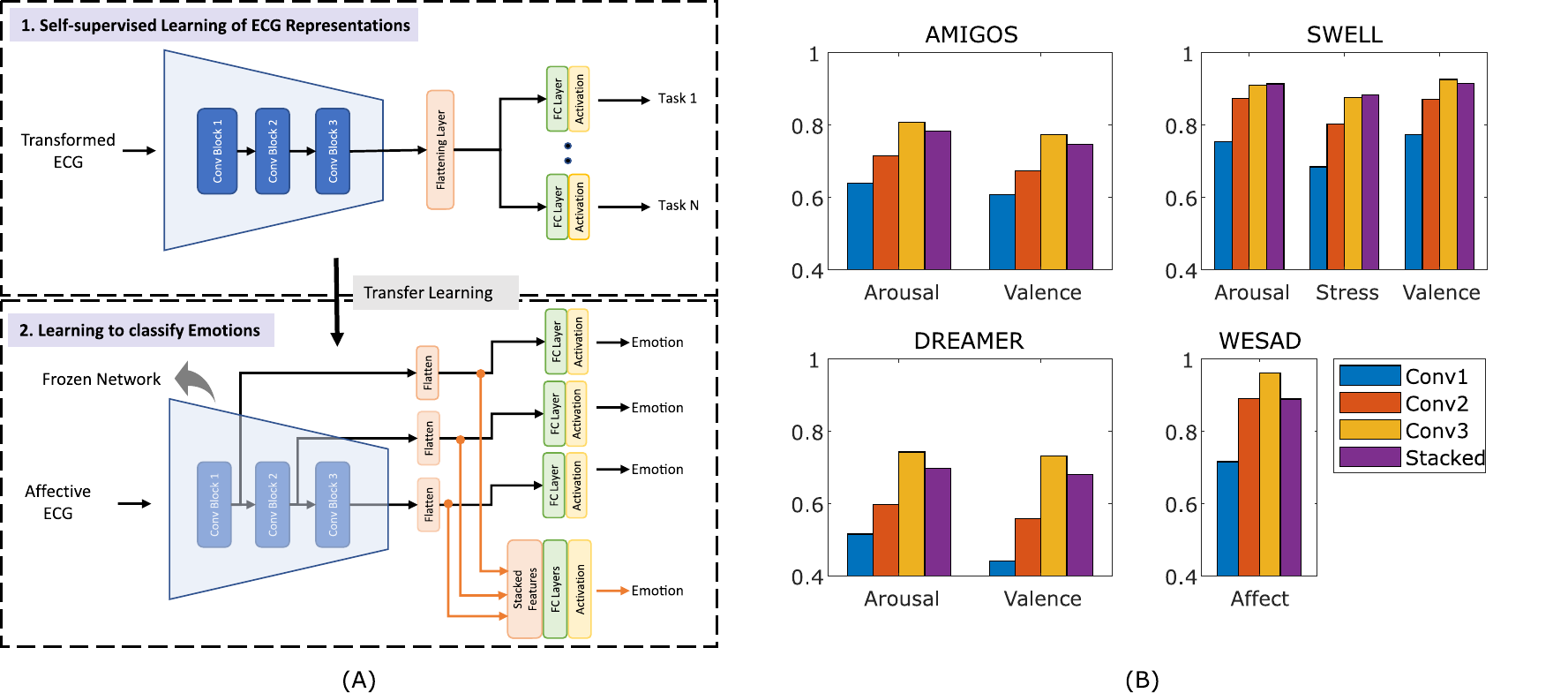}
    \caption{The feasibility of using embeddings obtained through the network at different depths for emotion recognition is investigated. In (A) the schematic of  the experiment is presented, while (B) presents the results.
    }
    \label{fig:layer_analysis}
\end{figure*}

\begin{figure}[t]
    \centering
    \includegraphics[width=0.8\columnwidth]{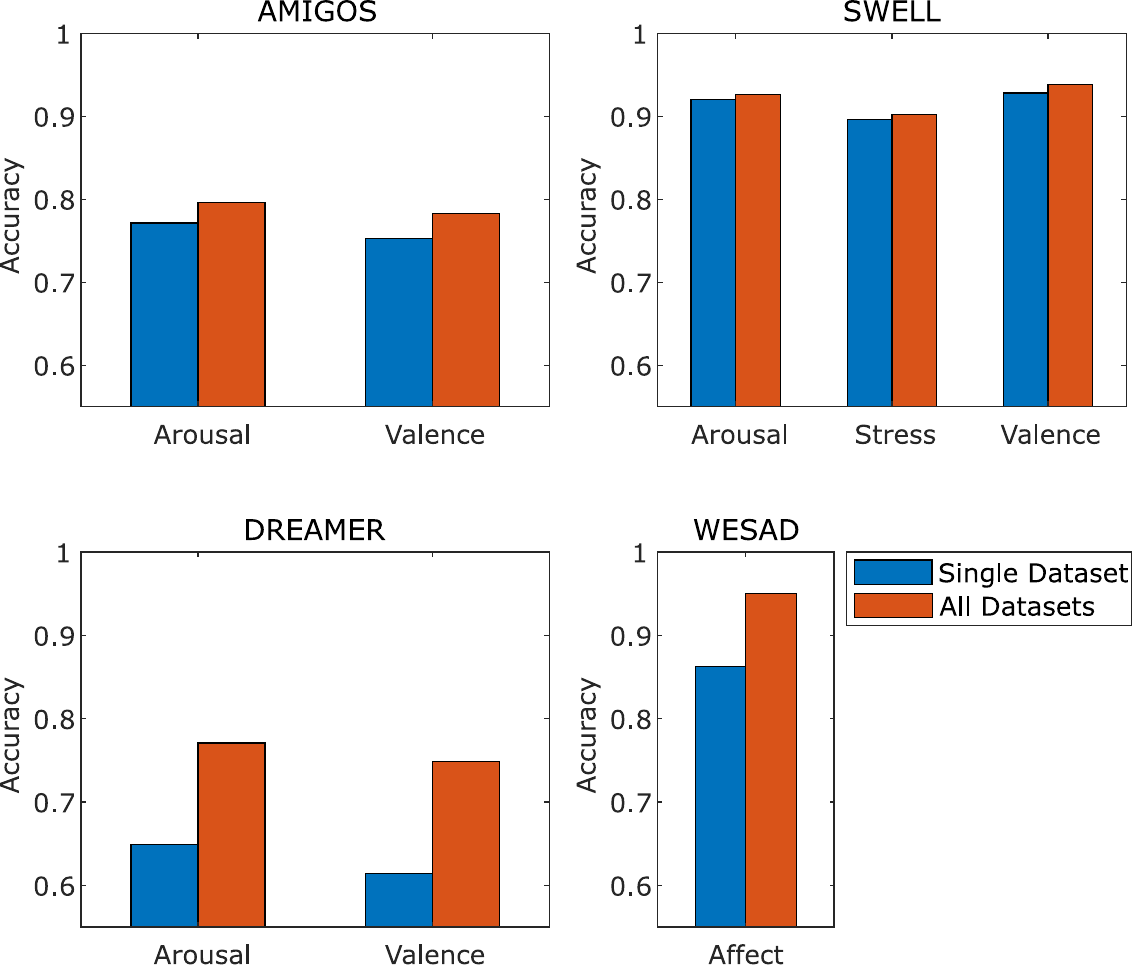}
    \caption{The impact of using multiple datasets with the self-supervised architecture is investigated and presented.}
    \label{fig:single_dataset}
\end{figure}


\subsection{Relationship Between Downstream and Pre-text Tasks} \label{sec:Relationship Between Downstream and Pre-text Tasks}
To further understand the relationship between the accuracy of classification of the pre-text tasks and downstream emotion recognition, we present this relationship in Figure \ref{fig:ervsstr}. In this figure, the transformation recognition accuracies are obtained based on varying degrees of task difficulty as discussed in Section \ref{Analysis of Individual Transformation Task}. Corresponding emotion recognition accuracies are then calculated for this analysis. Since similar patterns are noticed across different emotions and datasets, the average accuracies are utilized in the figure. For time-warping, there are two transformation parameters: number of segments and stretch factor. Therefore, first we keep the number of segments at optimum value according to the result achieved in Section \ref{Analysis of Individual Transformation Task}, and vary the other transformation parameter (stretch factor), and vice versa. Temporal inversion and negation transformation are not included here as there is no controllable parameters to vary the transformation recognition difficulty. 

The figure indicates that in the case of noise addition, scaling, and permutation, when the transformation recognition accuracy is poor, the emotion classification performance is also poor. This behaviour occurs when the transformation recognition tasks are too difficult. On the other hand, when the transformed signals are highly distorted compared to the original signal, the model reports high accuracy in transformation recognition, while poor performance is acquired for classifying emotions. This observation further confirms that for an optimum amount of difficulty in the pretext tasks, the self-supervised model shows the best performance for downstream tasks.

\subsection{Network Embeddings} \label{sec:Network Embeddings}
Here, we evaluate the quality of the learned representations vs. the depth of the self-supervised model. The representations learned with each conv-block are separately extracted and utilized to perform emotion recognition as shown in Figure \ref{fig:layer_analysis}(A). This analysis helps us understand whether representations at different depths of the network are informative and valuable for the downstream tasks.  
First, the learned representations from conv-block $1$, conv-block $2$, and conv-block $3$ are separately extracted and utilized to perform emotion recognition. As shown in Figure \ref{fig:layer_analysis}(B), significant improvement in performance is noticed when features are extracted from conv-block $3$ compared to conv-block $1$ and conv-block $2$. In addition to using the individual embeddings, we also stacked the three embeddings to create a new embedding, which we then used for the downstream tasks. However, the embedding obtained from conv-block $3$ still showed the best performance. 
This analysis indicates that learning representations obtained from the last convolutional layer is better and more generalizable for the downstream emotion recognition.

\subsection{Impact of Multiple of Datasets} \label{sec:Impact of Multiple of Datasets}
One of the major advantages of the self-supervised solution is that it allows for the self-supervised network to be trained using an aggregation of the existing datasets by removing the main barrier of having different types of output labels for each dataset.
To further analyze this aspect of the proposed solution, we investigate the impact of using all four datasets for training the network compared to when individual datasets are used. To perform this study, we use the four datasets separately to train the self-supervised signal transformation recognition network, which we then use for emotion recognition. Figure \ref{fig:single_dataset} shows that using all $4$ datasets combined, the model learns more generalized and robust features as evident by the better performance in classifying emotions compared to learning ECG representations from individual datasets. In the case of AMIGOS, DREAMER, and WESAD datasets, the performance improves significantly while marginal improvement is noticed for SWELL dataset.

\subsection{Limitations and Future Work} \label{sec:Limitations and Future Work}
In this paper we performed an extensive analysis of the proposed framework and highlight the benefits of the self-supervised approach over a fully-supervised technique. However, the limitations of the self-supervised solution should also be noted. We notice that the proposed solution performs poorly in subject-independent emotion recognition. We hypothesize the main possible reasons for this: \textit{1}) subject-\textit{invariant} features may require specific hand-crafted features or a combination of hand-crafted and deep learning features such as those used in \cite{pritam_acii} and \cite{siddharth2019utilizing} respectively, which might not be learnable solely with convolutions applied to the raw ECG; \textit{2}) Moreover, in order to properly normalize user-specific ECG features linked to factors such as genetics, physiology, health, and others, user-specific calibrations with respect to baseline sessions were carried out in \cite{pritam_acii}, which enabled subject-independent validation to perform well. We avoided such calibration steps in the pipeline since: \textit{a}) a calibration-free framework is generally more convenient and hence preferred, and \textit{b}) some of the datasets did not include the information required to perform such calibrations.

For future work, multi-modal emotion recognition will be explored, exploiting other modalities such as EEG available in some datasets. Furthermore, we will employ the proposed solution for other ECG-related domains such as arrhythmia detection, ECG-based activity recognition, and others. Lastly, further research towards utilization of the proposed architecture for cross-subject and cross-corpus schemes will be carried out.

\section{Conclusion} \label{conclusion_future_work}
In this work we present an ECG-based emotion recognition solution using self-supervised deep multi-task learning. To the best of our knowledge, this is the first time self-supervised learning is utilized to perform emotion recognition using ECG. Four public datasets AMIGOS, DREAMER, WESAD and SWELL are used in this study to perform emotion recognition. We set new state-of-the-art results for the four datasets in classification of arousal, valence, affective states, and stress. We show that the proposed approach significantly improves classification performance compared to a fully-supervised solution. We explore different self-supervised transformation recognition tasks for learning the ECG representations and analyze their impacts. Our analysis shows that for an optimum amount of difficulty in the pretext tasks, the network learns better ECG representations which can used for emotion classification. Our experiments also show that for learning of ECG representations, a multi-task CNN improves the performance compared to a single-task network. Lastly, our analysis on the use multiple datasets for training the self-supervised network shows that utilizing multiple datasets benefits the final downstream classification.

\bibliographystyle{unsrt}
\bibliography{ref.bib}

\end{document}